\documentstyle[12pt]{article}
\def\permil{\%\raise.10ex\hbox{$_{\scriptstyle 0}$}}
\def\nll{ \nonumber \\}
\def\la{\mathrel{\mathpalette\fun <}}
\def\ga{\mathrel{\mathpalette\fun >}}
\def\fun#1#2{\lower3.6pt\vbox{\baselineskip0pt\lineskip.9pt
  \ialign{$\mathsurround=0pt#1\hfil##\hfil$\crcr#2\crcr\sim\crcr}}}

\begin{document}

\title{\bf  Small Angle Bhabha Scattering for LEP
$^{\dagger}{}^{\ddagger}$}
\author{A.~Arbuzov$\;  ^{a}$
V.~Fadin$\;^{b}$
E.~Kuraev$\;   ^{a}$\\
L.~Lipatov$ \;  ^{c}$
N.~Merenkov$  \; ^{d}$
L.~Trentadue$ \;  ^{e}$}
\date{}

\maketitle

\begin{itemize}

\item[$^a$]
              Joint Institute for Nuclear Research,\\
Head Post Office, P.O. Box 79, Dubna 141980, Moscow region, Russia

\item[$^b$]
            Budker Institute for Nuclear Physics and Novosibirsk State
 University,\\
 630090, Novosibirsk, Russia

\item[$^c$]
        St.-Petersburg Institute of Nuclear Physics,\\
        188350  Gatchina, Leningrad region,  Russia

\item[$^d$]
             Physical-Technical Institute, 310108 Kharkov, Ukraine

\item[$^e$]
      Dipartimento di Fisica, Universit\'a di Parma, 43100 Parma \\
         INFN Sezione di Roma II, via della Ricerca Scientifica, 1, 00133 Roma,
 Italy\\
      present address: CERN, TH-Division, Geneva, Switzerland
\end{itemize}

\vspace{1cm}

\begin{abstract}
We present the results of our calculations to a one,
two, and three loop approximation of the e$^+$e$^-$$\rightarrow$e$^+$e$^-$
Bhabha
scattering cross-section at small angles. All terms  contributing to the
 radiatively
corrected
cross-section, within an accuracy of $\delta\sigma/ \sigma = 0.1 \% $,
are explicitely evaluated and presented in an analytic form.
$O(\alpha)$ and $O(\alpha^2)$ contributions are kept up to
next-to-leading
logarithmic accuracy, and $O(\alpha^3)$ terms are taken into account to
the leading logarithmic approximation.
We define an experimentally measurable cross-section  by integrating
the calculated distributions over a given range of final-state
energies and angles. The cross-sections for exclusive channels
as well as for the totally integrated distributions are also given.

\end{abstract}

\vfill

\footnoterule
\noindent
$^{\dagger} ${\footnotesize  Work supported by the $Istituto \;Nazionale\; di\;
Fisica\;Nucleare\;(INFN)$. INTAS grant 93-1867.}
$^{\ddagger}${\footnotesize Appeared as a contribution to the "Reports of
the Working Group on Precision Calculations for the Z Resonance" CERN 95-03,
March 1995.}
\vfill

\section{Introduction}

An accurate verification of the Standard
Model is one of the primary aims of LEP \cite{r1}.
While electroweak radiative corrections to the
$s$-channel annihilation process and to large angle Bhabha scattering
allow a direct extraction of the Standard Model parameters, small
angle Bhabha cross-section affects, as an overall
normalization condition, all observable cross-sections and
represents an equally unavoidable condition towards a precise
determination of the Standard Model parameters.
 The small angle Bhabha scattering process is used
to measure the luminosity of electron positron colliders. At LEP an
experimental accuracy on the luminosity of
\begin{eqnarray}\label{err}
      |\frac{\delta\sigma}{ \sigma}|< 0.001
\end{eqnarray}
will soon be reached \cite{r1a}. However, to obtain the total accuracy,
a systematic theoretical error must also be added. This precision calls for
an equally accurate theoretical expression for the Bhabha scattering
cross-section in order to extract the Standard Model
parameters from the observed distributions.
An accurate determination of the small angle Bhabha
cross-section and of the luminosity directly affects
the determination of
absolute cross-sections such as, for example, the determination of the
invisible width and of the number of massless neutrino species $N_{\nu}$
\cite{r111}.

In recent years a considerable attention has been
devoted to the Bhabha process \cite{r2,r3}. However, the
accuracy reached, is still inadequate. According to these
evaluations the theoretical estimates are still incomplete, moreover,
are  in disagreement with each other up
to
$ 0.5 \%$, far from the required theoretical and experimental accuracy
\cite{r1a}.

The process that  will be considered  in this work is that of Bhabha scattering
when electrons and positrons are emitted at small angles
with respect to the initial electron and positron directions. We have examined
the
radiative processes inclusively accompanying the main
e$^+$e$^-$$\rightarrow$e$^+$e$^-$ reaction at
high energies, when both the scattered electron and positron are tagged
within the counter aperture.

We assume that the center-of-mass energies are within the range of the LEP
 collider
$2\epsilon=\sqrt{s}=90-200\;$GeV
and the scattering angles are within the range $\theta \simeq 10-150\;$mrad.
We assume that the charged particle detectors have the following polar angle
cuts:

\begin{eqnarray}\label{angles}
\theta_1 < \theta_-=\widehat{\vec{p}_1 \vec{p}_{1'}\,} \equiv
\theta < \theta_3\;\;\;, \qquad
\theta_2 < \theta_+=\widehat{\vec{p}_2 \vec{p}_{2'}\,}<\theta_4\;\;\;,
\qquad 0.01\la \theta_i \la 0.1 \;\mbox{rad}\;,
\end{eqnarray}
where $\vec{p}_1 ,\, \vec{p}_{1'},\,\; (\vec{p}_2, \, \vec{p}_{2'}\,)$
represent the momenta of the initial and of the scattered electron (positron)
in
  the
center-of-mass frame.

In this paper we present the results of our calculations
of the Bhabha scattering
cross-section with an accuracy of $O(0.1\%)$. The squared matrix elements of
 the
various exclusive processes
inclusively contributing to the e$^+$e$^-$$\rightarrow$e$^+$e$^-$  reaction are
 integrated
in order to define  an experimentally measurable cross-section
according to suitable restrictions on the angles and energies of the
detected particles. The various contributions to the electron and positron
distributions, needed for the required accuracy, are presented
using analytical expressions.

In order to define the angular range of interest and the implications on
the required accuracy, let us first briefly discuss, in a general way, the
 angle-dependent
corrections to the cross-section.

We consider  e$^+$e$^-$  scattering at angles as defined in Eq. (\ref{angles}).
Within this region, if one expresses the cross-section by means of a series
expansion in terms of angles, the main contribution to the
cross-section
${d\sigma}/{d\theta^2}$
comes from the diagrams for the scattering amplitudes containing one  exchanged
 photon
in the $t$-channel. These diagrams, as it is well known, show a
singularity of the type $\theta^{-4}$ for $\theta \rightarrow 0$, e.g.
\begin{eqnarray}
 \frac{d\sigma}{d\theta^2} & \sim & \theta^{-4}\;\;\;.\nonumber
\end{eqnarray}

Let us now estimate the correction of order $\theta^2$ to this contribution.
If
\begin{eqnarray}
 \frac{d\sigma}{d\theta^2} & \sim & \theta^{-4}(1+c_1\theta^2)\;\;\; ,
\end{eqnarray}
then, after integration over  $\theta^2$  in the angular range as
Eq. (\ref{angles}),
one obtains:

\begin{eqnarray}
\int \limits_{\theta^2_{\mbox{\tiny min}}}^{\theta^2_{\mbox{\tiny max}}}
\frac{d\sigma}{d\theta^2}\;d\theta^2
& \sim & \theta^{-2}_{\mbox{\tiny min}}(1+c_1\theta^2_{\mbox{\tiny min}}
\ln \frac{\theta^2_{\mbox{\tiny max}}}{\theta^2_{\mbox{\tiny min}}}).
\end{eqnarray}
 We see that, for  $\theta_{\mbox{\tiny min}} = 50 \;$mrad and
$\theta_{\mbox{\tiny max}} = 150\;$mrad (we have taken
the case where the $\theta^2$  corrections are maximal), the relative
contribution of the $\theta^2$ terms is about  $5 \mbox{x} 10^{-3} c_1.$
Therefore,
the terms of relative order $\theta^2$  must only be kept in the
Born cross-section where the coefficient $c_1$ is not small.
In higher orders of the
perturbative expansion the coefficient
$c_1$  contains at least one factor ${\alpha}/{\pi}$ and therefore these
terms can safely be omitted. This implies that, within our accuracy,
only radiative corrections from the scattering type diagrams contribute.
Furthermore, one should take into account only diagrams with
one photon exchanged in the $t$-channel, since,
according to the generalized eikonal representation, the large logarithmic
terms from the diagrams with the multi-photon exchange are cancelled.

Having, as a final goal for the experimental cross-section, the relative
 accuracy
as in Eq. (\ref{err}), and by taking into account that the minimal value of
the squared momentum transfer
$Q^2 = 2\epsilon^2 (1-\cos\theta)$ in the region (\ref{angles}) is of the
order of $1\;\mbox{GeV}^2$, we  may omit  in the following also the terms
appearing in the radiative corrections of the type
${m^2}/{Q^2}$ with $m$ equal to the electron $(m_e)$, or
the muon $(m_{\mu})$ mass.

The contents of this paper can be outlined as follows. In Section 2
we discuss the Born cross-section $d\sigma^B$, taking the
$Z^0$ boson exchange into account, and compute
the corrections to it caused by the virtual and  real soft photon
emission.
We present the results, as discussed above,
in the form of an expansion in terms of the scattering angle $\theta$.
We then define an experimentally measurable cross-section,
$\sigma_{\mbox{\tiny exp}}$,
which is obtained by tagging the scattered electron and positron within a
 suitable
range of polar angles and energies. We introduce the ratio
$\Sigma={\sigma_{\mbox{\tiny exp}}}/{\sigma_0}$ by  normalizing
$\sigma_{\mbox{\tiny exp}}$ with
 respect to the cross-section $\sigma_0={4 \pi \alpha^2}/
{\epsilon^2\theta^2_1}$.
In Section 2, by using a simplified version of the differential cross-section
for the small angle scattering, we discuss the contribution to
$\sigma_{\mbox{\tiny exp}}$ from the single bremsstrahlung process.
In Section 3  we find all  corrections of $O(\alpha^2)$ to
$\sigma_{\mbox{\tiny exp}}$ caused by two virtual and
real photon emissions. In Section 4 we consider $O(\alpha^2)$ corrections
caused by e$^+$e$^-$ pair emission.
In Section 5 we derive the leading logarithmic corrections to the
$\alpha^3$ order by using  the structure function method.
In Section 6 we estimate the contributions of the neglected terms.
Finally, in Section 7 we give the results obtained in terms of the ratio
 $\Sigma$
as functions of the experimental parameters.

A more detailed derivation of these results will be presented elsewhere
\cite{r33}.

\section{Born and one-loop soft and virtual corrections}

  The Born cross-section for Bhabha scattering within the Standard Model is
well known \cite{r4}:
\begin{equation}\label{born}
 \frac{d\sigma^B }{d\Omega }\, = \,\frac{\alpha^2 }{8s}\;\; \bigl\{
4B_1 +(1-c)^2B_2 +(1+c)^2 B_3 \bigr\}\;\;,
\end{equation}
where
\begin{eqnarray*}
B_1 &=& (\frac{s}{t})^2 |1+(g^2_v-g^2_a)\xi |^2 \; ,
\qquad B_2= | 1+(g^2_v-g^2_a) \chi |^2 \;,       \\
B_3 &=&\frac{1}{2} | 1+\frac{s}{t}+(g_v+g_a)^2(\frac{s}{t} \xi +\chi)|^2+
\frac{1}{2} | 1+\frac{s}{t}+(g_v-g_a)^2(\frac{s}{t} \xi +\chi)|^2 \; ,     \\
\chi &=& \frac{\Lambda s}{s-m^2_z+iM_Z \Gamma _Z}\;, \qquad \xi = \frac{\Lambda
t}{t-M^2_Z} \;,
\\ \Lambda &=& \frac{G_FM^2_Z}{2\sqrt 2 \pi \alpha}=(\sin 2\theta
_w)^{-2}\;,\;\;\;
 g_a=-\frac{1}{2}\;g_v=-\frac{1}{2} (1-4 \sin^2\theta_w)\;,
   \\
s &=& (p_1 +p_2)^2 =4\varepsilon ^2\;, \qquad t=-Q^2=(p_1
-p_{1'})^2=-\frac{1}{2}\;s\;(1-c)\;,
\\ c &=& \cos\theta \;,\qquad \theta =\widehat{\vec p_1 \vec p_{1'}}\;.
\end{eqnarray*}
Here $\theta_w$ is the Weinberg angle.
In the small angle limit
\begin{eqnarray*}
c=\cos\theta=1-\frac{\theta^2}{2}+\frac{\theta^4}{24}-
\frac {\theta^6}{720}+ ...\;\;\;.
\end{eqnarray*}
Expanding the result (5) we have
\begin{eqnarray}\label{tborn}
     \frac{d\sigma^B}{\theta d\theta}=\frac{8 \pi \alpha^2}{\epsilon^2
\theta^4}
 \;(1-\frac{\theta^2}{2}+\frac{9}{40}\; \theta^4+\delta_{\mbox{\tiny weak}}),
\end{eqnarray}
where $\epsilon={\sqrt {s}}/{2}$ is the electron or positron initial
energy. The contribution from weak interactions
$\delta_{\mbox{\tiny weak}}$, connected with diagrams with
$Z^0$ boson
exchange, is given by the expression:
\begin{eqnarray}\label{dweak}
\delta_{\mbox{\tiny weak}}=2g_v^2\;
 \xi-\frac{\theta^2}{4}\;(g_v^2+g_a^2)\;Re\chi
+\frac{\theta^4}{32}\;(g_v^4+g_a^4+6g_v^2g_a^2)\;|\chi|^2\;\;.
\end{eqnarray}

{}From Eq. (7) one can see that the contribution $c^w_1$  of the weak
correction
$\delta_{\mbox{\tiny weak}}$  into the coefficient $c_1$ in Eq. (3) is
\begin{equation}\label{cw}
c^w_1\; \la\; 2g_v^2+\frac{(g_v^2+g_a^2)}{4}\; \frac{M_Z}{\Gamma_Z}
+\theta^2_{\mbox{\tiny
 max}}\;\frac{(g_v^4+g_a^4+6g_v^2g_a^2)}{32}\;\frac{M^2_Z}{\Gamma^2_Z}
\sim 1\;\;.
\end{equation}
According to our previous discussion after Eq. (4) this means that the
 contribution connected
with  the $Z^0$ boson exchange diagrams does not exceed $0.3 \%$.
We shall therefore neglect the radiative corrections to weak contributions,
since they could contribute at most with terms $\la 10^{-4}$.

In the pure QED case one-loop radiative corrections to the Bhabha
cross-section were calculated a long time ago \cite{r5}.
Taking into account the contribution coming from the emission of soft
photons with energy less than a  given finite threshold $\Delta \epsilon$
as well, one obtains for
${d \sigma^{(1)}_{\mbox{\tiny QED}}}/{dc}$ in one-loop approximation
\begin{eqnarray}\label{qed1}
\frac{d \sigma^{(1)}_{\mbox{\tiny QED}}}{dc}=\frac
{d \sigma^{B}_{\mbox{\tiny QED}}}{dc}\;(1+
\delta_{\mbox{\tiny virtual}}+ \delta_{\mbox{\tiny soft}})\;\;,
\end{eqnarray}
where $d \sigma^{B}_{\mbox{\tiny QED}}$ is the Born cross-section in the pure
QED case ( i.e. it is equal to $d \sigma^{B}$ with \\ $g_a =g_v = 0$ ) and
\begin{eqnarray}
\delta_{ \mbox{\tiny virtual}}+ \delta_{\mbox{\tiny soft}}
& = & 2\; \frac{\alpha}{\pi} \; \{2 \;[1- \ln\;
(\frac{4\epsilon^2}{m^2}) +2\ln \,
ctg \;\frac { \theta}{2}] \ln
\frac {\epsilon} {\Delta \epsilon} +\int_{\cos^2 \frac{\theta}{2}}^{\sin^2
\frac{\theta}{2}} \frac {dx}{x} \ln(1-x) \nonumber\\
& &  -\frac{23}{9}+\frac{11}{6} \; \ln\; (\frac{4\epsilon^2}{m^2}) \}
+\frac{\alpha}{\pi} \;\frac{1}{(3+ c^2)^2}\;[ \frac{ \pi^2}{3}\;(2 c^4 -3 c^3
-15 c)
\nonumber \\ & &
+2\;(2 c^4- 3 c^3 + 9 c^2+ 3 c+ 21)\;\ln^2 \sin \frac{ \theta}{2}
\nonumber \\ & &
-4\;(c^4 + c^2 - 2 c)\;
\ln^2 \cos \frac{ \theta} {2}- 4\;( c^3 + 4 c^2+ 5 c+ 6)\;\ln^2 tg \frac
{\theta} {2}
 \;\;\;\;\;\;\;
\nonumber \\ & &
+ \frac{2}{3}\;(11 c^3+33 c^2+21 c+111)\;\ln\sin \frac{ \theta}{2}
+2\;(c^3-3 c^2+7 c-5)\;\ln \cos \frac{ \theta}{2}
\nonumber \\ & &  \nonumber
+2\;(c^3+3 c^2+3 c+9) \;\delta_t-2\;( c^3+3 c)(1-c) \;\delta_s]\;\;.
\end{eqnarray}
The value $\delta_t$  ($\delta_s$)   is defined by the contributions to the
 photon
vacuum polarization function $\Pi(t)$  [$\Pi(s)$]  as follows:
\begin{eqnarray}\label{pi}
\Pi(t)=\frac{\alpha}{\pi} \;(\delta_t+\frac{1}{3}L-\frac{5}{9})
+\frac{1}{4}\;(\frac{\alpha}{\pi})^2 L\;\;,
\end{eqnarray}
where
\begin{eqnarray}\label{l}
L=\ln\frac{Q^2}{m^2}\;\;,\;\;  Q^2=-t=2\epsilon^2 (1-c)\;\;,
\end{eqnarray}
and we have taken into account the leading logarithmic part of the two-loop
corrections to the
polarization operator. In the Standard Model $\delta_t$ contains
contributions of muons, tau-leptons, W-bosons, and hadrons:
\begin{eqnarray}
\delta_t= \delta_t^{\mu} + \delta_t^{\tau} + \delta_t^{W}
+\delta_t^H\;, \qquad \delta_s=\delta_t\;(Q^2\rightarrow-s)\;,
\end{eqnarray}
and the first three contributions are
theoretically calculable:
\begin{eqnarray}
&& \delta_t^{\mu}=\frac{1}{3}\ln\frac{Q^2}{m_{\mu}^2}-\frac{5}{9}\;,
\nonumber\\
 && \delta_t^{\tau}= \frac{1}{2}\;v_{\tau}\;(1-\frac{1}{3}v^2_{\tau})
\;\ln\frac{v_{\tau}+1}{v_{\tau}-1}+\frac{1}{3}\;v^2_{\tau}
- \frac{8}{9}\;, \qquad
v_{\tau}=(1+\frac{4m_{\tau}^2}{Q^2})^{\frac{1}{2}}\;;\\
 && \delta_t^{w}= \frac{1}{4}\;v_{w}\;(v^2_{w}-4)
\;\ln\frac{v_{w}+1}{v_{w}-1}-\frac{1}{2}\;v^2_{w}
+\frac{11}{6}\;, \qquad
v_w=(1+\frac{4m_{w}^2}{Q^2})^{\frac{1}{2}}\;\;.   \nonumber
\end{eqnarray}
The contribution of the hadrons $\delta_t^H$ can be expressed through
the experimentally measurable cross-section of the
e$^+$e$^-$ annihilation
\cite{r6}.

In the limit of small scattering angles
we can present Eq. (9) in the following form:
\begin{eqnarray}\label{qeds1}
&& \frac{d \sigma^{(1)}_{\mbox{\tiny QED}}}{dc}=\frac
{d \sigma^{B}_{\mbox{\tiny QED}}}{dc}\;
[1-\Pi(t)]^{-2}\;(1+\delta)\;,
\\ \nonumber &&
\delta=2 \frac{\alpha}{\pi}\;[2(1-L)\ln\frac{1}{\Delta}+\frac{3}{2}L-2]+
\frac{\alpha}{\pi}\;\theta^2\;\Delta_{\theta}
+\frac{\alpha}{\pi}\;\theta^2\; \ln\Delta\;,
\\ \nonumber &&
\Delta_{\theta}=\frac{3}{16}l^2+\frac{7}{12}l-\frac{19}{18}
+\frac{1}{4}\;(\delta_t-\delta_s)\;,
\\ \nonumber &&
\Delta=\frac{\Delta\epsilon}{\epsilon},  \qquad l=\ln\frac{Q^2}{s}\simeq
 \ln \;\frac {\theta^2}{4}\;.
\end{eqnarray}
This representation gives us a possibility to verify explicitly that
the terms of a relative order $\theta^2$ in the radiative corrections are
small.
Taking into account that the large contribution proportional to $\ln\Delta$
disappears when we add the cross-section for the hard emission, one can
verify once more that such terms can be
neglected. Therefore in higher orders
we will omit the annihilation diagrams as
well as multiple-photon exchange diagrams in the scattering channel.
The second simplification is justified by the generalized eikonal
representation for the amplitudes at small scattering angles \cite{r7}.

Let us introduce now the dimensionless quantity

\begin{eqnarray}
\Sigma=\frac{Q_1^2\;\sigma_{\mbox{\tiny exp}}}{4\pi\alpha^2}\;\;,\nonumber
\end{eqnarray}
with $ Q_1^2=\epsilon^2 \theta_1^2$ where  $\sigma_{\mbox{\tiny exp}}$
 represents the
experimentally observable cross-section:

\begin{eqnarray}\label{cross}
\Sigma=\frac{Q_1^2}{4\pi\alpha^2}\int\limits_{}^{}\,dx_1\int\limits_{}^{}\,dx_2
\;\theta(x_1x_2-x_c)\int\limits_{}^{}\,d^2q_1\;\theta_1^c\int\limits_{}^{}\,
d^2q_2
\;\theta_2^c\;\;\;\\ \nonumber
\frac{d\sigma^{\mbox{\tiny e}^+\mbox{\tiny e}^-\rightarrow \mbox{\tiny e}^
+(\vec{\mbox{\tiny q}}_2,\mbox{\tiny x}_2)\;
\mbox{\tiny e}^-(\vec{\mbox{\tiny q}}_1,\mbox{\tiny x}_1)+\mbox{\tiny
 X}}}{dx_1d^2q_1dx_2d^2q_2}\;\;,\;\;\;\;\;\;\;
\;\;\;\;\;\;\;\;\;\;\;\;\;\;\;\;\;\;\;
\end{eqnarray}
 where $x_{1,2}$ and $\vec{q}_{1,2}$  are, respectively, the energy fractions
 and
the transverse components of the electron and
positron  momenta in the final state, $sx_c$ is the experimental cut-off
on their squared invariant mass $sx_1x_2$, and the functions
$\theta_i^c$ which take into account the angular cuts  in Eq. (2) are
defined as:

\begin{eqnarray}\label{cuts}
\theta_1^c=\theta(\theta_3-\frac{|\vec{q}_1|}{x_1\epsilon})\;\;
\theta(\frac{|\vec{q}_1|}{x_1\epsilon}-\theta_1)\;\;, \qquad
\theta_2^c=\theta(\theta_4-\frac{|\vec{q}_2|}{x_2\epsilon})\;\;
\theta(\frac{|\vec{q}_2|}{x_2\epsilon}-\theta_2)\;\;.
\end{eqnarray}
We restrict ourselves further to the symmetrical case only:

\begin{eqnarray}
\theta_2=\theta_1\;,\quad\;\; \theta_4=\theta_3\;,\quad\;\;
\rho=\frac{\theta_3}{\theta_1} > 1\;.
\end{eqnarray}

Let us define $\Sigma $ as a sum of exclusive contributions:

\begin{equation}\label{sigma}
\Sigma=\Sigma_0+\Sigma^{\gamma}+\Sigma^{2\gamma}
+\Sigma^{\mbox{\tiny e}^+ \mbox{\tiny e}^-}+\Sigma^{3\gamma}+\Sigma^
{\mbox{\tiny e}^+\mbox{\tiny e}^-\gamma}\;\;,
\end{equation}
where $ \Sigma_0$ stands for a modified Born contribution, $\Sigma^{\gamma}$,
$\Sigma^{2\gamma}$, and $\Sigma^{3\gamma}$ stand
for the contributions of one, two, and three photon emissions (both  real and
virtual),
$\Sigma^{\mbox{\tiny e}^+\mbox{\tiny e}^-}$ and
$\Sigma^{\mbox{\tiny e}^+\mbox{\tiny e}^-\gamma}$ represent the
emission of virtual or real (soft and hard) pairs without
and with the accompaining real or virtual photon.

By integrating Eq. (\ref{tborn}) with the use of the full  propagator for the
$t$-channel
photon, which takes into account the growth of the electric charge at small
distances, we obtain:
\begin{equation}\label{sigma0}
\Sigma_0=\theta_1^2\int\limits_{\theta_1^2}^{\theta_2^2}
\frac{d\theta^2}{\theta^
4}
(1-\Pi(t))^{-2}+\Sigma_W+\Sigma_\theta\;\;,
\end{equation}
 where
\begin{eqnarray}
\Sigma_W=\theta_1^2\int\limits_{\theta_1^2}^{\theta_2^2}\frac{d\theta^2}
{\theta^4}\;\delta_{\mbox{\tiny weak}}\;\;, \nonumber
\end{eqnarray}
is the correction due to the weak interactions and the term

\begin{eqnarray}
\Sigma_{\theta}=\theta_1^2\int\limits_{1}^{\rho^2}\frac{dz}{z}\;
[1-\Pi(-zQ_1^2)]^{-2}\;(-\frac{1}{2}+z\;\theta_1^2\;\frac{9}{40})\;\;,
\nonumber
\end{eqnarray}
comes from the expansion of the Born cross-section in Eq. (\ref{born})
in powers of $\theta^2$.

The one-loop contribution $\Sigma^{\gamma}$ comes from one-photon emission
(real and virtual). By adding to Eq. (14) the cross-section for hard
emission calculated using a simplified version of the differential
 cross-section for small angle scattering \cite{r9} we obtain:

\begin{eqnarray}\label{sigmagam}
&& \Sigma^{\gamma}=\frac{\alpha}{\pi}\int\limits_{1}^{\rho^2}\,
\frac{dz}{z^2} [1-\Pi(-zQ_1^2)]^{-2}\{\int\limits_{x_c}^{1}
dx\;[(L_z-1)P(x)
\\ \nonumber && \qquad \;
[1+\theta(x^2\rho^2-z)]+\frac {1+x^2}{1-x}k(x,z)] - 1\}\;\;,
\end{eqnarray}
where

\begin{eqnarray}\label{p}
P(x)=(\frac{1+x^2}{1-x})_+=\lim_{ \Delta \to 0 }\;
\bigl\{ \frac{1+x^2}{1-x}\;
\theta(1-x-\Delta)+(\frac{3}{2}+2\ln\Delta)\;\delta(1-x) \bigr\}\;\;,
\end{eqnarray}
is the non-singlet splitting kernel and

\begin{eqnarray}\label{k}
k(x,z)= \frac{(1-x)^2}{1+x^2} \;[1+\theta(x^2\rho^2-z)]
+ \; L_1+\theta(x^2\rho^2-z)\;L_2
\\ \nonumber
+ \theta(z-x^2\rho^2) L_3\;\;.\;\;\;\;\;\;\;\;\;\;\;\;\;\;\;\;\;\;\;\;\;\;\;\;
\end{eqnarray}

Here $L_z=\ln \frac {z\;Q_1^2}{m^2}$ and

\begin{eqnarray}\label{l123}
 L_1=\ln|\frac{x^2(z-1)(\rho^2-z)}{(x-z)(x\rho^2-z)}|\;\;, \quad
 L_2=\ln|\frac{(z-x^2)(x^2\rho^2-z)}{x^2(x-z)(x\rho^2-z)}|\;\;,
\\ \nonumber
 L_3=\ln|\frac{(z-x^2)(x\rho^2-z)}{(x-z)(x^2\rho^2-z)}|\;\;.\;\;\;
\;\;\;\;\;\;\;\;\;\;\;\;\;\;
\end{eqnarray}

\section{Two-photon emission}

Let us now consider the corrections of the order of  $\alpha^2$. They
come from the two-photon emission as well as from the pair production,
real and virtual. The virtual and soft real photon corrections can be
obtained by using the results of Refs. \cite{r9}-\cite{r11}.

The corresponding contributions to $\Sigma$ are:

\begin{eqnarray}\label{sigmasv}
\Sigma_{S+V}^{\gamma\gamma}=\Sigma_{VV}
+\Sigma_{VS}+\Sigma_{SS}=(\frac{\alpha}{\pi})^2
\int\limits_{1}^{\rho^2}\,dz\;z^{-2}
(1-\Pi(-zQ_1^2))^{-2}R_{S+V}^{\gamma\gamma}\;\;.
\end{eqnarray}
It is convenient to present the $R_{S+V}^{\gamma\gamma}$ in the following
way:
\begin{eqnarray}\label{rsv}
R_{S+V}^{\gamma\gamma}=r_{S+V}^{(\gamma\gamma)}+r_{S+V(\gamma\gamma)}+
r_{S+V\gamma}^{\gamma}\;\;\;\;\;\;\;\;\;\;\;\;\;\;\;\;\;\;\;\;\;\;\;\;\;\;
\\ \nonumber
r_{S+V}^{(\gamma\gamma)}= r_{S+V(\gamma\gamma)}= L_z^2 (2 \;\ln^2\; \Delta
+ 3 \;\ln
\;\Delta + \frac{9}{8})
+ L_z (- 4 \;\ln^2 \Delta - 7 \;\ln\; \Delta +3 \xi_3- \frac{3}{2} \xi_2 -
 \frac{45}{16})
\\ \nonumber
r_{S+V\gamma}^{\gamma}=4[(L_z-1)\;\ln\;\Delta+{3\over 4}L_z-1]^2\;\;.
\;\;\;\;\;\;\;\;\;\;\;\;\;\;\;\;\;\;\;\;\;\;\;\;\;\;\;\;
\end{eqnarray}
Here the quantity
$ \Delta= \delta \epsilon / \epsilon\; <<1 $
is the maximal energy fraction carried by a soft photon.

The single hard photon radiation can be accompanied
by real soft or virtual photons. It is useful to separate the cases
of photons emitted by the same electron or positron or by both of them.

\begin{eqnarray}\label{dsigmahsv}
d\sigma|_{H,S+V}=d\sigma^{H(S+V)}+d\sigma_{H(S+V)}+d\sigma^{H}_{(S+V)}+
d\sigma_{H}^{(S+V)}\;\;.
\\ \nonumber && \qquad
\end{eqnarray}

In the case where one of the fermions emits the hard real photon and
another interacts with a virtual or soft real photon, we find:

\begin{eqnarray}\label{sigmahsv}
\Sigma^H_{(S+V)}+\Sigma_H^{(S+V)}=2\frac{\alpha}{\pi}
\int\limits_{x_c}^{1-\Delta} \,dx
\frac{1+x^2}{1-x} \int\limits_{1}^{\rho^2}\,
dzz^{-2}[1-\Pi(-zQ_1^2)]^{-2}\;\;\;\;\;\;\;\;\;\;
\\ \nonumber
\;\;\{ [1 +\theta(x^2\rho^2-z)]\;
(L_z-1)+k(x,z) \}({\alpha \over \pi})[(L_z-1)\;\ln\;\Delta
+{3 \over 4}L_z-1]\;\;.
\\ \nonumber
\end{eqnarray}

%where $\Sigma^H= \frac{\alpha}{\pi} \int\limits_{x_c}^{1-\Delta} \,dx
%\frac{1+x^2}{1-x} \int\limits_{1}^{\rho^2}\,
%dzz^{-2}(1-\Pi(-zQ_1^2))^{-2}\;\cdot\;\{ [1 +\theta(x^2\rho^2-z)]\;
%(L-1)+k(x,z) \}$.

 A more complicated expression arises when both  photons interact with
  the same fermion.
In this case the cross-section can be expressed in terms
of the Compton tensor with a 'heavy photon' \cite{r12}.
 We will consider below the case of  the photon emission from
the electron. An equal contribution arises from the positron activity.
For the small angle detection of the final fermions
we have \cite{r4}:

\begin{eqnarray}\label{hardvs}
\Sigma^{H(S+V)} &=& \Sigma_{H(S+V)} \\ \nonumber
&=& \frac{1}{2}({\alpha \over \pi})^2
\int_1^{\rho^2}\frac{dz}{z^2(1-\Pi(-zQ_1^2))^{2}}\int_{x_c}^
{1-\Delta}\frac{dx(1
 +x^2)}{1-x}\;L_z
\;\{(2\;\ln\;\Delta-\ln \;x+\frac{3}{2})\;\;\;\;
\\ \nonumber
&&[(L_z-1)(1+\theta)+k(x,z)]
+\frac{1}{2}\;\ln^2x+(1+\theta)[-2+\ln \;x-2\;\ln\;\Delta]\;\;\;\;\;\;\;\;\;\;
\\ \nonumber
&&+(1-\theta)[\frac{1}{2} L_z\; \ln\; x+2\;\ln\; \Delta \;\ln\; x-\ln\; x
\; \ln(1-x)\;\;\;\;
\\ \nonumber
&&-\ln^2x+\int_0^{1-x} \frac{dt}{t}\; \ln\;(1-t)
-\frac{x(1-x)+4x\;\ln \;x}{2(1+x^2)}]-\frac{(1-x)^2}{2(1+x^2)}\}\;\;,\;\;\;
\;\;\;\;\;\;\;\;
\end{eqnarray}
where $\theta=\theta(x^2\rho^2-z)$.

Let us consider now the contribution from the emission of two
hard real photons. One can distinguish two cases: double photon
bremsstrahlung,  a) in opposite directions along electron and positron
momenta, and b) in the same direction along electron or positron momenta.

The differential cross-section in the first case can be obtained by
using the factorization of cross-sections in the impact parameter
space \cite{r12}.
It takes the following form \cite{r9,r4}:

\begin{eqnarray}\label{ff}
\Sigma_H^H=\frac{1}{4}(\frac{\alpha}{\pi})^2\int\limits_{0}^{\infty}
\,dz\;z^{-2}[1-\Pi(-zQ_1^2)]^{-2}\int\limits_{x_c}^{1-\Delta}\,dx_1
\int\limits_{\frac{x_c}{x_1}}^{1-\Delta}\,dx_2\;
\\ \nonumber
\frac{1+x_1^2}{1-x_1}\;\frac{1+x_2^2}{1-x_2}\;\Phi(x_1,z)\Phi(x_2,z)\;\;,
\;\;\;\;\;\;\;\;\;\;\;\;\;\;\;\;\;\;\;\;\;\;\;\;\;\;\;
\end{eqnarray}
 where

\begin{eqnarray}\label{phi}
\Phi(x,z)=(L_z-1)[\theta(z-1)\theta(\rho^2-z)+\theta(z-x^2)
\theta(\rho^2x^2-z)]\;\;\;\;\;\;\;\;\;\;\;\;
\nonumber \\
+L_3[-\theta(x^2-z)+\theta(z-x^2\rho^2)]
+[L_2+\frac{(1-x)^2}{1+x^2}]\theta(z-x^2)\theta(x^2\rho^2-z)
\;\;\;\;\;\;\;\;
\\ \nonumber
+[L_1+\frac{(1-x)^2}{1+x^2}]\theta(z-1)\theta(\rho^2-z)+[\theta(1-z)
-\theta(z-\rho^2)]\;\ln\;|
\frac{(z-x)(\rho^2-z)}{(x\rho^2-z)(z-1)}|\;\;\;\;\;\;\;.
\end{eqnarray}

Let us now turn to  the double hard photon emission in the
same direction, and the hard e$^+$e$^-$ pair production.
We distinguish the cases of the
collinear and semi-collinear kinematics of the final particles. In the first
case all particles produced move in the cones within the polar angles
 $\theta_0<<
\theta_1$ centred along the charged particle momenta (final or initial).
The region
corresponding to the case when both photons are radiated
outside these cones does not contain any large logarithmic contribution.
In the semi-collinear region only one of the particles produced moves inside
the cones, whilst the other moves outside them.

In the totally inclusive
cross-section the
dependence on
the auxiliary parameter $\theta_0$ disappears, and the total contribution
has the form:

\begin{eqnarray}\label{hh}
\Sigma^{HH}=\Sigma_{HH}=\frac{1}{4}(\frac{\alpha}{\pi})^2
\int\limits_{1}^{\rho^2}\,dz\,z^{-2}[1-\Pi(-zQ_1^2)]^{-2}\int\limits_{x_c}
^{1-2\Delta}\,dx\int\limits_{\Delta}^{1-x-\Delta}\,dx_1
\nonumber \\
\frac{I^{HH}L_z}
{x_1(1-x-x_1)(1-x_1)^2}\;\;,\;\;\;\;\;\;\;\;\;\;\;\;\;\;\;\;\;\;\;\;\;\;\;
\\ \nonumber
I^{HH}=A\;\theta(x^2\rho^2-z)+B+C\;\theta[(1-x_1)^2\rho^2-z]\;\;\;\;\;\;\;
\;\;\;.
\end{eqnarray}
Here
\begin{eqnarray}
A=\gamma\beta(\frac{L_z}{2}+\ln \frac{(\rho^2 x^2-z)^2}
{x^2(\rho^2 x (1-x_1)-z)^2})+(x^2+(1-x_1)^4)\ln\frac{(1-x_1)^2(1-x-x_1)}{xx_1}
+\gamma_A\;\;,
\nonumber \\ \nonumber
B=\gamma\beta(\frac{L_z}{2}+\ln |\frac{x^2(z-1)(\rho^2-z)(z-x^2)
(z-(1-x_1)^2)^2(\rho^2x(1-x_1)-z)^2}
{(\rho^2 x^2-z)(z-(1-x_1))^2(\rho^2(1-x_1)^2-z)^2(z-x(1-x_1))^2}| )
\;\;\;\;\;\;\;\;\;\;\;\;\;\;\;\;\;
\\ \nonumber
+(x^2+(1-x_1)^4)\ln\frac{(1-x_1)^2x_1}{x(1-x-x_1)}
+\delta_B\;\;,\;\;\;\;\;\;\;\;
\;\;\;\;\;\;\;\;\;\;\;\;\;\;\;\;
\nonumber \\
C=\gamma\beta (L_z+2\ln| \frac{x(\rho^2 (1-x_1)^2-z)^2}{(1-x_1)^2
(\rho^2 x (1-x_1)-z)(\rho^2(1-x_1)-z)}|)
-2(1-x_1)\beta-2x(1-x_1)\gamma\;\;,\;\;\;\;\;\;\;\;\nonumber
\\
\gamma=1+(1-x_1)^2\;\;,\quad \beta=x^2+(1-x_1)^2\;\;,\;\;\;\;\;\;\;\;
\;\;\;\;\;\;\;\;\;\;\;\;\;\;\;\;\;\;\;\;\;\;\;\;\;\;\;\;\;\;\;\;\;\;\;\;\;
\\
\gamma_A=xx_1(1-x-x_1)-x_1^2(1-x-x_1)^2 -2(1-x_1)\beta \;\;,\;\;\;\;\;\;\;\;
\;\;\;\;\;\;\;\;\;\;\;\;\;\;\;\;\;\;\;\;\;
\nonumber
\\  \delta_B=xx_1(1-x-x_1)-x_1^2(1-x-x_1)^2 -2x(1-x_1)\gamma.
\;\;\;\;\;\;\;\;\;\;\;\;\;\;\;\;\;\;\;\;\;\;\;\;\;\;\;\;\;\;\;
 \nonumber
\end{eqnarray}

One can verify \cite{r33} that the combinations

\begin{eqnarray}\label{rgg}
({\alpha \over \pi})^2 \int_1^{\rho^2}\frac{dz}{z^2[1-\Pi(-zQ_1^2)]^{2}}
r^{\gamma\gamma}_{S+V}+\Sigma^{H(S+V)}+\Sigma^{HH}\;\;,\;\;
 \nonumber \\
({\alpha \over \pi})^2 \int_1^{\rho^2}\frac{dz}{z^2[1-\Pi(-zQ_1^2)]^{2}}
r^{\gamma}_{S+V\gamma}+\Sigma^{H}_{S+V}+\Sigma_H^{S+V}+\Sigma^{H}_{H}\;\;,\;\;
\end{eqnarray}
do not depend on $\Delta$ for $\Delta\rightarrow 0$.

The total expression $ \Sigma^{2\gamma} $ which describes the
contribution to Eq. (\ref{sigma}) from all  (real and virtual) two-photon
 emissions is determined by the expressions in Eqs. (\ref{sigmasv}),
 (\ref{rsv}),
(\ref{sigmahsv}), (\ref{hardvs}), (\ref{ff}), and (\ref{hh}).

 Furthermore, it does not depend on the auxiliary parameter
$\Delta$ and can be written as follows:

\begin{eqnarray}\label{s2g}
\Sigma^{2\gamma}=\Sigma_{S+V}^{\gamma\gamma}+2\;\Sigma^{H(V+S)}
+\Sigma^H_{(V+S)}+\Sigma_H^{(V+S)}+\Sigma_H^H+2\Sigma^{HH}\\ \nonumber
=\Sigma^{\gamma\gamma}+\;\Sigma_{\gamma}^{\gamma}+
(\frac{\alpha}{\pi})^2{\cal{L}}\phi^{\gamma\gamma}\;\;,\;\;\;\;\;\;\;\;\;\;\;\;
{\cal{L}}=\ln\frac{\epsilon^2\Theta_1^2}{m^2}\;\;.\;\;\;\;\;\;\;\;
\end{eqnarray}

The leading contributions $ \Sigma^{\gamma\gamma},
\Sigma_{\gamma}^{\gamma} $ have the form:

\begin{eqnarray}\label{s2gamma}
&& \Sigma^{\gamma\gamma}=\frac{1}{2}(\frac{\alpha}{\pi})^2\int\limits_{1}^
{\rho^2}L_z^2\,dz\;z^{-2}
[1-\Pi(-Q_1^2z)]^{-2}\int\limits_{x_c}^{1}\,dx\;\{\frac{1}{2}P^{(2)}(x)\;[
\;\theta(x^2\rho^2-z)+1]
\nonumber \\ &&  \qquad
 + \int\limits_{x}^{1}\,\frac{dt}{t}
P(t)\;P(\frac{x}{t})\;\theta(t^2\rho^2-z)\}\;\;,
\end{eqnarray}
where

\begin{eqnarray}\label{p2}
P^{(2)}(x)=\int\limits_{x}^{1}\,\frac{dt}{t}P(t)\;P(\frac{x}{t})=
\lim_{\Delta \to 0}
 \{\;[(2\ln\Delta+\frac{3}{2})^2-4\xi_2]\;\delta(1-x)\;\;\;\;\;\;\;\;
\;\;\;\;\;\;\;\;\;\;\;\;\;\;\;\;\;\;\;\;\;\;\;\;\;\;\;\;
\\ \nll
%&& \qquad
+\;2\;[\frac{1+x^2}{1-x}(2\ln(1-x)-\ln x + \frac{3}{2})+\frac{1}{2}(1+x)\ln x
- 1+x]\;\theta(1-x-\Delta)\}\;\;,\;\;\;\;\;\;\;\;\nonumber
\end{eqnarray}
and

\begin{eqnarray}\label{sgg}
\Sigma_{\gamma}^{\gamma}&=&\frac{1}{4}(\frac{\alpha}{\pi})^2
\int\limits_{0}^{\infty}\,L_z^2 dz\;z^{-2}
[1-\Pi(-Q_1^2z)]^{-2}\int\limits_{x_c}^{1}\,dx_1
\int\limits_{{x_c}/{x_1}}^{1}\,dx_2
P(x_1)P(x_2)[\theta(z-1)\theta(\rho^2-z) \nonumber
\\
%&&  \qquad
&&
+\theta(z-x_1^2)\theta(x_1^2\rho^2-z)]
[\theta(z-1)\theta(\rho^2-z)
+\theta(z-x_2^2)\theta(x_2^2\rho^2-z)]\;\;.\;\;\;\;\;\;\;\;\;\;\;\;
\end{eqnarray}

We see that the leading contributions to $\Sigma^{2\gamma}$ can be expressed
in terms of kernels for the evolution equation for structure functions.
 The function $\phi^{\gamma\gamma}$ in  the expression (\ref{s2g}) collects the
next-to-leading contributions which cannot be obtained by the
structure functions method \cite{r15,r101}. It has a form given
explicitely in Ref. \cite{r33}.

\section{Pair production}
In a similar way we consider also pair production.
The  corrections due to virtual  e$^+$e$^-$  pairs  can  be extracted from
Ref. \cite{r10}.  Using the expression for soft pair production
cross-section \cite{r11} one obtains for the
contribution of the virtual and soft real e$^+$e$^-$ pairs to $\Sigma$
the following result:

\begin{eqnarray}\label{sigmasvee}
&& \Sigma_{S+V}^{\mbox{\tiny e}^+\mbox{\tiny
 e}^-}=(\frac{\alpha}{\pi})^2\int\limits_{1}^{\rho^2}\,
dzz^{-2} [1-\Pi(-zQ_1^2)]^{-2}R_{S+V}^{\mbox{\tiny e}^+\mbox{\tiny e}^-}\;\;,
\\
&& R_{S+V}^{\mbox{\tiny e}^+\mbox{\tiny e}^-}=L_z^2(\frac{2}{3}\ln\Delta
+\frac{1}{2}) + L_z(-\frac{17}{6}+\frac{4}{3}\ln^2\Delta \nonumber
\\ \nonumber && \qquad
 -\frac{20}{9}\ln\Delta-\frac{4}{3}\xi_2)+O(1)\;\;.
\\ \nonumber && \qquad
\end{eqnarray}
 In  this expression the quantity
$ \Delta= \delta \epsilon / \epsilon\; <<1 $
is the maximal energy fraction carried by a soft pair.

Here we have taken into account only e$^+$e$^-$ pairs.
An order of magnitude of the pair production  correction is
less than  $0.5\%$. A  rough estimate of the muon  pair contribution gives
less than  $0.05\%$ since $\ln{Q^2}/{m^2}\sim
3\ln{Q^2}/{m_{\mu}^2}$.
Contributions of pion and tau-lepton pairs give even smaller corrections.
Therefore, within the $0.1\%$  accuracy, we can omit any pair
except e$^+$e$^-$ .

Let us consider now the hard pair production.
In this case we can restrict ourselves only to the collinear region
where the produced pair moves in the small cones within the polar angles
$\theta_0$ around the fermion momenta.
Indeed, the non-leading semi-collinear
region will give contributions of one order of magnitude smaller
than the leading ones and therefore within the required accuracy one
can neglect them \cite{r4}.

The pair production contributions to $\Sigma $ appear from
two regions with  the pair components moving respectively along initial
and  scattered electrons and analogously for positrons.
This separation allows us to carry out the integration over the energy
fraction of the pair
components. The resulting contribution to $\Sigma$ has,
with both directions included, the form :

\begin{eqnarray}\label{sighee}
&& \Sigma_H^{\mbox{\tiny e}^+\mbox{\tiny
 e}^-}=\frac{1}{4}(\frac{\alpha}{\pi})^2\int\limits_{1}^
{\rho^2}dz\;z^{-2}[1-\Pi(-zQ_1^2)]^{-2}
\int\limits_{x_c}^{1-\Delta}dx\{R_0(x)\,[{\cal{L}}^2\,[1+\theta(x^2\rho^2-z)]
\;\;\;\;\;\;\;
\\ \nonumber && \qquad
+\;4{\cal{L}}\;\ln \;x]+2 \;\theta(x\rho-1)\;\theta(x^2\rho^2-z)\;
 {\cal{L}}\;f(x)+
2 f_1(x)\;{\cal{L}}\}\;\;,
\end{eqnarray}
where we put the auxiliary parameter $\theta_0^2$ equal to
$\theta^2=z\theta_1^2$, and introduce the following notations:

\begin{eqnarray}
 R_0(x)&=&\frac{2}{3}\;\frac{1+x^2}{1-x}+
\frac{1-x}{3x}(4+7x+4x^2)+2(1+x)\ln x\;,
\\ \nonumber
f(x)&=&-\frac{131}{9}+\frac{136}{9}x-\frac{2}{3}x^2 -\frac{4}{3x}
-\frac{20}{9(1-x)}+\frac{2}{3}(-4x^2-5x+1+\frac{4}{x}
\\ \nonumber
%&& \qquad
&&
+\frac{4}{1-x})\ln\;(1-x)
+\frac{1}{3}(8x^2+5x-7-\frac{13}{1-x})\ln x-2\frac{1}{1-x}\ln^2x
%\\  && \qquad
\\  &&
+\frac{4x^2}{1-x}\int\limits_{0}^{1-x}
dy\frac{\ln\;(1-y)}{y}+2(1+x)[\xi_2+\ln
 x\ln\;(1-x)+\int\limits_{0}^{x}dy\frac{\ln\;(1-y)}{y}]\;\;.
\\ \nonumber
f_1(x)&=&-\frac{116}{9}+\frac{151}{9}x
+\frac{2}{3x}+\frac{4x^2}{3}
-\frac{20}{9(1-x)}+\frac{1}{3}[8x^2-10x-10+\frac{5}{1-x}]\ln x
\\ \nonumber && \qquad
+\frac{2}{3}[-4x^2-5x+1+\frac{4}{x}
+\frac{4}{1-x}]\ln(1-x)-(1+x)\ln^2x\;\;\;\;\;\;\;\;\;\;\;\;\;\;\;
%\\ \qquad
\\
&&
+2(x+1)[\xi_2+\int\limits_{0}^{x}
dy\frac{\ln\;(1-y)}{y}+\ln x\ln\;(1-x)]
-\frac{4}{1-x}\int\limits_{0}^{1-x}dy\frac{\ln\;(1-y)}{y}\;\;,
\end{eqnarray}
where $f_1(x)=-xf(\frac{1}{x})$.

The total contribution of virtual, soft, and hard pairs
does not depend on the auxiliary parameter $\Delta$ and can be written as
follows:

\begin{eqnarray}\label{sigmaeeg}
&& \Sigma^{\mbox{\tiny e}^+\mbox{\tiny e}^-}=
\Sigma_{S+V}^{\mbox{\tiny e}^+\mbox{\tiny e}^-}+\Sigma_H^{\mbox{\tiny e}^
+\mbox{\tiny e}^-}=
\frac{1}{4}(\frac{\alpha}{\pi})^2{\cal{L}}^2\int\limits_{1}^{\rho^2}\,
dzz^{-2}[1-\Pi(-zQ_1^2)]^{-2}
\\ \nonumber && \qquad
\int\limits_{x_c}^{1}\,dxR(x)
[\theta(x^2\rho^2-z)+1]+(\frac{\alpha}{\pi})^2
{\cal{L}}\phi^{\mbox{\tiny e}^+\mbox{\tiny e}^-}\;\;,
\end{eqnarray}
where

\begin{eqnarray}
R(x)=2\;(1+x)\ln x+\frac{1-x}{3x}(4+7x+4x^2)+\frac{2}{3}P(x)\;\;.
\end{eqnarray}

The explicit expression for $\phi^{\mbox{\tiny e}^+\mbox{\tiny e}^-}$
which collects the non-leading
terms from Eqs. (\ref{sigmasvee}) and (\ref{sighee}) is :

\begin{eqnarray}
\phi^{\mbox{\tiny e}^+\mbox{\tiny
 e}^-}=\frac{1}{2}\int\limits_{1}^{\rho^2}\,dzz^{-2}
\int\limits_{x_c}^{1}\,dx[\theta(x\rho-1)\theta(x^2\rho^2-z)\bar f
+\bar f_1-\frac{17}{6}-\frac{4}{3}\xi_2]\;\;,\;\;\;\;\;\;\;\;\;\;\;\;
\end{eqnarray}
where
\begin{eqnarray}
\bar f=f+\frac{20}{9}[\frac{1}{1-x}-(\frac{1}{1-x})_+]+\frac{8}{3}
(-\frac{\ln(1-x)}{1-x}+(\frac{\ln(1-x)}{1-x})_+)\;\;.\;\;\;\;\;\;\;\;\;\;\;\;
\end{eqnarray}

The same definition applies to $\bar f_1 $. The regularization corresponding
to the $(\;\;\;)_+$ prescription
is as in Ref. \cite{r15}.

\section{$O(\alpha^3)$ corrections}

In order to evaluate the leading logarithmic contribution
represented by terms of the type $(\alpha {\cal {L}})^3$
we use the iteration of the
master equation obtained in Refs. \cite{r15,r101}.

To simplify the analytical expressions we adopt here a realistic assumption
about the smallness of the threshold for the detection of the hard
subprocess energy. By neglecting the terms of the order of:

\begin{eqnarray}
x_c\;(\frac{\alpha}{\pi} {\cal{L}})^3\sim 10^{-4}\;\;,
\end{eqnarray}
one should consider only the emission by the initial particles.
Three photon (virtual and real) contribution to $\Sigma$ have the form:

\begin{eqnarray}\label{3g}
&& \Sigma^{3\gamma}=\frac{1}{4}\;(\frac{\alpha}{\pi} {\cal{L}})^3\;
\int\limits_{1}^{\rho^2}\, dz\;
z^{-2}\int\limits_{x_c}^{1}\,dx_1
\int\limits_{x_c}^{1}\,dx_2\;\theta(x_1x_2-x_c)\;
[\frac{1}{6}\;\delta(1-x_2)\;P^{(3)}(x_1)
\\ \nonumber && \qquad
\theta(x_1\rho-1)\theta(x_1^2\rho^2-z)
+\frac{1}{2}P^{(2)}(x_1)P(x_2)\theta_1\theta_2]\;[1+O(x_c^3)]\;\;,
\end{eqnarray}
where $P(x)$  and  $P^{(2)}(x)$ are given by Eqs. (\ref{p}) and
(\ref{p2}) correspondingly,

\begin{eqnarray}
&& \theta_1\theta_2 = \theta(z-\frac{x_2^2}{x_1^2})\;
\theta(\rho^2 \frac{x_2^2}{x_1^2}-z)\;\;,
\nonumber \\ &&
P^{(3)}(x)=\delta(1-x)\;\Delta_t+\theta(1-x-\Delta)\;\theta_t\;\;,
\nonumber\\  &&
\Delta_t=48 \;[\frac{1}{2}\xi_3 -\frac{1}{2}\xi_2(\ln \Delta + \frac{3}{2})
+\frac{1}{6}(\ln \Delta + \frac{3}{2})^3 ]\;\;,
\\ \nonumber &&
\theta_t= 48 \;\{\frac{1}{2}\;\frac{1+x^2}{1-x}
\;[\frac{9}{32}-\frac{1}{2}\xi_2
+\frac{3}{4}\ln(1-x)-\frac{3}{8}\ln x+\frac{1}{12}\ln^2(1-x)
\\ \nonumber && \qquad
+\frac{1}{12} \ln^2x-\frac{1}{2}\ln x\ln (1-x)]
+\frac{1}{8}(1+x)\ln x\ln(1-x)-\frac{1}{4}
(1-x)\ln(1-x)
\\ \nonumber && \qquad
+\frac{1}{32}(5-3x)\ln x -\frac{1}{16}(1-x)-\frac{1}{32}(1+x)\ln^2x
-\frac{1}{8}(1+x)\int\limits_{0}^{1-x}\,dy\frac{\ln(1-y)}{y}\}\;\;.
\end{eqnarray}

The contribution to $\Sigma$ of the pair production accompanied by the
photon emission when both, pair and photons, can be real and virtual is given
below (with respect to Ref. \cite{r15} we include also the non-singlet
mechanism
of the pair production):

\begin{eqnarray}\label{eeg}
&& \Sigma^{\mbox{\tiny e}^+\mbox{\tiny
 e}^-\gamma}=\frac{1}{4}(\frac{\alpha}{\pi}
{\cal{L}})^3\int\limits_{1}^{\rho^2}\,
dz\;z^{-2}\int\limits_{x_c}^{1}\,dx_1\int\limits_{x_c}^{1}\,dx_2
\;\theta(x_1x_2-x_c)\;
\nonumber \\ \nonumber && \qquad
\{\frac{1}{3}[R^P(x_1)-\frac{1}{3}R^s(x_1)]\;\delta(1-x_2)
\theta(x_1^2\rho^2-z)+\frac{1}{2}\;P(x_2)R(x_1)
\;\theta_1\theta_2\}\;[1+O(x_c^3)]\;\;,
\end{eqnarray}
 where

\begin{eqnarray}
&& R(x)=R^s(x)+\frac{2}{3}P(x)\;\;, \qquad
R^s(x)=\frac{1-x}{3x}(4+7x+4x^2)+2(1+x)\ln x\;\;,
\\ \nonumber &&
R^P(x)=R^s(x)(\frac{3}{2}+2\ln(1-x))+(1+x)(-\ln^2x-4
\int\limits_{0}^{1-x}\,dy\frac{\ln(1-y)}{y})
\\ \nonumber && \qquad
+\frac{1}{3}(-9-3x+8x^2)\ln x+
\frac{2}{3}(-\frac{3}{x}-8+8x+3x^2)\;.
\end{eqnarray}

The total expression for $\Sigma$ in Eq. (\ref{sigma}) is the sum of
the contributions given in Eqs. (\ref{sigma0}), (\ref{sigmagam}), (\ref{s2g}),
 (\ref{sigmaeeg}), (\ref{3g})
and (\ref{eeg}).
The quantity $ \Sigma $ is a function of the parameters $x_c,\rho$, and
$Q_1^2$.

\section{Estimates of neglected terms}

Let us now estimate the terms not taken into account
here in accordance with the required accuracy:

\vskip 10.0pt
a)Weak radiative corrections:

\begin{eqnarray}
\Sigma^{\mbox{\tiny w.r.c.}}\sim\frac{\alpha Q_1^2}{\pi M_z^2}\la 10^{-5}\;\;.
\end{eqnarray}

\vskip 10.0pt
b) Electromagnetic corrections  to weak contributions,
 including interference terms :
\vskip 10.0pt

\begin{eqnarray}
\Sigma_{W}^{\mbox{\tiny h.o.}}\sim \delta_{\mbox{\tiny
weak}}|_{\theta=\theta_1}
{\alpha\over{\pi}}{\cal{L}}\la 10^{-4}\;\;.
\end{eqnarray}
Here $\delta_{\mbox{\tiny weak}}$ is given by Eq. (7).
\vskip 10.0pt
c) Radiative corrections to the annihilation mechanism,  including its
 interference with the scattering
mechanism :

\begin{eqnarray}
\Sigma_{\mbox{\tiny st}}^{\mbox{\tiny r.c.}}\sim \theta_1^2
\frac{\alpha}{\pi}{\cal{L}} \la 10^{-4}\;\;.
\end{eqnarray}

Our explicit expressions
for $\Sigma^{\gamma}$,  without annihilation terms,
coincide numerically with the results obtained
at the same order in Ref. \cite{r16} by using exact matrix elements.
\vskip 10.0pt
d) The interference between photon emissions by electron and positron:

\begin{eqnarray}
\Sigma_{\mbox{\tiny int}}\sim \theta_1^2 \frac{\alpha}{\pi}\la 10^{-5}\;\;.
\end{eqnarray}
This contribution is connected with terms violating the eikonal form \cite{r7}
 in the
expression:

\begin{eqnarray}
A(s,t)=A_0(s,t)e^{i\phi(t)}+O(\frac{\alpha t}{\pi s})\;\;.
\end{eqnarray}

\vskip 10.0pt
e) The interference terms between one- and two-photon
mechanisms of pair production, including the effect
final particles identity:

\begin{eqnarray}
\Sigma_{\mbox{\tiny int}}^{\mbox{\tiny pairs}}\sim
(\frac{\alpha}{\pi})^2\la 10^{-5}\;\;.
\end{eqnarray}
\vskip 10.0pt
f) The semi-collinear mechanism of pair production gives a contribution
which contains a small factor ${\cal{L}}^{-1}$
with respect to the collinear terms, and is numerically small:
$\Sigma_{\mbox{\tiny SC}}^{\mbox{\tiny pair}}\la 10^{-4}$. A more accurate
estimate should be interesting.

\vskip 10.0pt
g) The creation of heavy pairs ($ \mu\mu,\tau\tau,\pi\pi,...$) is
at least one order of magnitude smaller than the corresponding
contribution due to the light particle production and is therefore
not essential.

\vskip 10.0pt
h) Higher-order corrections, including soft and collinear multi-photon
contributions, can be safely
neglected since they only give contributions of the
type $({\alpha\;{\cal{L}}}/{\pi})^n$ for $n\geq4$.

\section{Results and their discussion}

The total cross-section for the $\mbox{\tiny e}^+\mbox{\tiny e}^-$ distribution
 is:

\begin{eqnarray}
\sigma=\frac{4\pi\alpha^2}{Q_1^2}\Sigma\;\;,
\end{eqnarray}
where $\Sigma$ is given by Eq. (\ref{sigma}).

Let us  define $\Sigma_0^0$ to be equal to $\Sigma_0|_{\Pi=0}$ [see Eq. (19)],
which corresponds to the Born cross-section obtained by switching
out the vacuum polarization contribution $\Pi(t)$ defined in Eq. (10).
We obtain:

\begin{eqnarray}\label{deli0}
\sigma=\frac{4\pi\alpha^2}{Q_1^2}\Sigma_0^0\;(1+\delta_0+\delta^{\gamma}
+\delta^{2\gamma}+\delta^{\mbox{\tiny e}^+\mbox{\tiny e}^-}
+\delta^{3\gamma}+\delta^{\mbox{\tiny e}^+\mbox{\tiny e}^-\gamma})\;\;,
\end{eqnarray}
 where

\begin{eqnarray}
\Sigma_0^0=\Sigma_0|_{\Pi=0}=1-\rho^{-2}+\Sigma_W+\Sigma_{\theta}|_{\Pi=0}
\simeq 1-\rho^{-2}\;\;,
\end{eqnarray}
and

\begin{eqnarray}
\delta_0=\frac{\Sigma_0-\Sigma_0^0}{\Sigma_0^0}\;;\;\;\;
\delta^{\gamma}=\frac{\Sigma^{\gamma}}{\Sigma_0^0}\;;\;\;\;
\delta^{2\gamma}=\frac{\Sigma^{2\gamma}}{\Sigma_0^0}\;;\;\;\;
\cdots\;\;\;.
\end{eqnarray}

In the Tables below we give the values of $\delta_{\beta}$ as well as
their sum as functions of $x_c$ for the  values of the parameters
$\theta_1=1.61^0$, and $\theta_2=2.80^0$ defining the Narrow-Narrow $(NN)$
case, and for $\theta_1=1.5^0$, and $\theta_2=3.15^0$,
defining the  Wide-Wide $(WW)$ case as in Ref. \cite{r18}.

Each of these contributions to $\Sigma$ has a sign which can be  changed
as a result of the interplay between real and virtual corrections.
The cross-section corresponding to  the Born diagrams for producing
a real particle is always positive, whereas
the sign of the radiative corrections depends on the order of
perturbation theory for the virtual corrections:  at odd orders it is
negative, and at even orders it is positive.
When the aperture of the counters
is small the compensation between real and virtual corrections
is not complete. In the limiting case of zero aperture only the virtual
contributions remain giving a negative result. As a consequence we
see
 that the
radiative corrections for the $NN$ case are larger in absolute value than
ones in the $WW$ case. These corrections depend on $\theta_1,x_c$, and $\rho$.
When $x_c\rightarrow 1$ the corrections increase in absolute value.

\section{Acknowledgements}
One of us (L.T.) would like to thank M. Dallavalle, B. Pietrzyk and T. Pullia
for several useful discussions at various stages of the work.
Two of us (E.K. and N.M.) would like to thank the INFN Laboratori Nazionali
di Frascati and, in particular, Miss Stefania Pelliccioni, the Theory Group at
 the
Dipartimento di Fisica of the
Universit\'a di Roma 'Tor Vergata' and the Dipartimento di Fisica of the
Universit\'a di Parma for their hospitality at several
stages during the preparation of this work.

%-----------------------------------------------------
%\section{References and comments}

\newpage
\section*{Tables}

\begin{tabular}{|c|c|c|c|c|c|c|c|}
\hline
$x_c$ & $\delta_0 $ & $\delta^{\gamma} $
&$\delta^{2\gamma} $ & $\delta^{\mbox{\tiny e}^+\mbox{\tiny e}^-} $ &
 $\delta^{3\gamma} $
&$\delta^{\mbox{\tiny e}^+\mbox{\tiny e}^-\gamma} $&$\sum \delta^i $ \\ \hline
\multicolumn{8}{|c|}{NN~~~~$\rho=1.74$} \\ \hline
0.2 & 0.413E-1& -0.922E-1& 0.506E-2& -0.750E-3&
     -0.561E-03& 0.911E-4& -0.471E-1 \\
0.3 & 0.413E-1& -0.962E-1& 0.508E-2& -0.633E-3&
     -0.514E-03& 0.767E-4& -0.509E-1 \\
0.4 & 0.413E-1& -0.101E+0& 0.452E-2& -0.635E-3&
     -0.483E-03& 0.669E-4& -0.562E-1 \\
0.5 & 0.413E-1& -0.108E+0& 0.406E-2& -0.672E-3&
     -0.453E-03& 0.589E-4& -0.637E-1 \\
0.6 & 0.413E-1& -0.119E+0& 0.385E-2& -0.733E-3&
     -0.414E-03& 0.517E-4& -0.749E-1 \\
0.7 & 0.413E-1& -0.138E+0& 0.348E-2& -0.823E-3&
     -0.350E-03& 0.456E-4& -0.943E-1 \\
0.8 & 0.413E-1& -0.174E+0& 0.397E-2& -0.964E-3&
     -0.249E-03& 0.442E-4& -0.130E+0 \\
\hline
\multicolumn{8}{|c|}{WW~~~~$\rho=2.10$} \\ \hline
0.2 & 0.409E-1& -0.728E-1& 0.276E-2& -0.579E-3&
     -0.406E-03& 0.598E-4& -0.301E-1 \\
0.3 & 0.409E-1& -0.771E-1& 0.290E-2& -0.491E-3&
     -0.377E-03& 0.506E-4& -0.341E-1 \\
0.4 & 0.409E-1& -0.826E-1& 0.257E-2& -0.502E-3&
     -0.356E-03& 0.442E-4& -0.399E-1 \\
0.5 & 0.409E-1& -0.902E-1& 0.223E-2& -0.543E-3&
     -0.333E-03& 0.387E-4& -0.479E-1 \\
0.6 & 0.409E-1& -0.104E+0& 0.200E-2& -0.606E-3&
     -0.297E-03& 0.338E-4& -0.620E-1 \\
0.7 & 0.409E-1& -0.127E+0& 0.207E-2& -0.697E-3&
     -0.238E-03& 0.304E-4& -0.849E-1 \\
0.8 & 0.409E-1& -0.167E+0& 0.303E-2& -0.838E-3&
     -0.146E-03& 0.320E-4& -0.124E+0 \\
\hline
\end{tabular}

\vspace{.5cm}
{\bf Table 1:}  The values of $\delta^i$
in Eq. (\ref{deli0}) for  $\sqrt{s}=M_Z=91.161$ GeV,
$\sin^2\Theta_W=0.2283$, $\Gamma_Z=2.4857$ GeV. The
'NN counter' corresponds to $\rho=1.74$ and
the 'WW counter' to $\rho=2.10$ as a function of $x_c$.

\vspace{2.5cm}

\begin{tabular}{|c|c|c|c|c|c|c|c|}
\hline
$x_c$    & 0.2 & 0.3 & 0.4 & 0.5  & 0.6  &0.7 &0.8  \\
\hline
$R_{NN}$ & 0.592 & 0.627 & 0.708 & 0.770 & 0.817 & 0.763 & 0.498 \\
\hline
$R_{WW}$ & 0.653& 0.740& 0.883 & 1.020 & 1.120 & 0.994 & 0.500 \\
\hline
\end{tabular}

\vspace{.5cm}
\hspace{.5cm} {\bf Table 2:}  Values of $R_{NN}$ and $R_{WW}$
where $R$ represents the ratio of non-leading with respect to leading
contributions and is defined as
$R=(\frac{\alpha}{\pi})^2 \frac {{\cal {L}} \phi^{2\gamma}}{
\Sigma^{2\gamma}}$
for  $\sqrt{s}=M_Z=91.161$ GeV,
$\sin^2\Theta_W=0.2283$, $\Gamma_Z=2.4857$ GeV. The
'NN counter' corresponds to $\rho=1.74$ and
the 'WW counter' to $\rho=2.10$ as above as a function of $x_c$.

\def\beq#1{\begin{equation}\label{#1}}
\def\beeq#1{\begin{eqnarray}\label{#1}}
\def\eeq{\end{equation}}
\def\eeeq{\end{eqnarray}}

%\appendix
%\chapter{}
%\input{appendixd}
%\chapter{}
%\input{appendixf}


\begin{thebibliography}{99}
\bibitem{r1}
     G. Altarelli, Lectures given at the E. Majorana Summer Scool, Erice,
       Italy, July 1993, CERN-TH.7072/93.

\bibitem{r1a}
     The {\it LEP Collaboration: ALEPH, DELPHI, L3, and OPAL,} and the {\it LEP
Electroweak Working Group}, CERN/PPE/93-157, 26 August 1993;
\\
B. Pietrzyk, High Precision Measurements of the Luminosity at LEP, Invited talk
 at the 'Radiative Corrections Status and Outlook' Conference, Gatlinburg, TN,
 July 1994,
LAPP-EXP-94.18, October 1994.

\bibitem{r111}
       $Neutrino\;Counting$ in $Z\;Physics\;at\;LEP$,
       G. Barbiellini et al. , L. Trentadue (conv.),
       G. Altarelli, R. Kleiss and C. Verzegnassi eds., CERN Report 89-08
 (1989).

\bibitem{r2}
       W. Beenakker, F.A. Berends and S.C. van der Marck, Nucl. Phys. B355
       (1991) 281;
\\     M. Cacciari, A. Deandrea, G. Montagna, O. Nicrosini and L. Trentadue,
 Phys.
       Lett. B271 (1991) 431;
\\     M. Caffo, H. Czyz and E. Remiddi, DFUB91-01 , April 1991;
\\     K.S. Bjorkenvoll, G. Faldt, and P. Osland, Nucl. Phys. B386 (1992)
       280, 303;
\\
       W. Beenakker and B. Pietrzyk, Phys. Lett. B296 (1992) 241 and
       CERN-TH.6760 (1992).

\bibitem{r3}
       S. Jadach et al.,  Phys. Rev. D47 (1993) 3733;
\\     S. Jadach, E. Richter-Was, B. F. L. Ward and
       Z. Was Phys. Lett. B260 (1991) 438; ibidem 268 (1991) 253;
\\     S. Jadach, M. Skrzypek and B.F L. Ward, Phys. Lett. B257 (1991) 173.

\bibitem{r33} V.S. Fadin, E.A. Kuraev, L.N. Lipatov, N.P. Merenkov and
L. Trentadue, in preparation.

\bibitem{r4}
       R. Budny,  Phys. Lett. 55B (1975) 227;
\\     D. Bardin, W. Hollik and T. Riemann, MPI-PAE/Pth, 32/90, PHE90-9,
       June, 1990.
\\     M. Boehm, A. Denner and W. Hollik, Nucl. Phys. B304 (1988) 687.

\bibitem{r5}
       R.V. Polovin, ZhETF 31 (1956) 449 ;
\\     F.A. Redhead, Proc. Roy. Soc. 220 (1953) 219;
\\     F.A. Berends et al. Nucl. Phys. B68 (1974) 541;
\\     F.A. Berends et al. Nucl. Phys. B228 (1983) 537.

\bibitem{r6}
       F. Jegerlehner, Prog. Part. Nucl. Phys. 27 (1991) 1,
       and references therein.

\bibitem{r7}
       E.A. Kuraev, L.N. Lipatov and N. P. Merenkov, Phys. Lett. 47B (1973) 33;
       Preprint 46 LNPI, 1973;
\\     H. Cheng and T. T. Wu, Phys. Rev. 187 (1969) 1868;
\\     V.S. Fadin, E.A. Kuraev, L.N. Lipatov, N.P. Merenkov and L. Trentadue,
       Yad. Fiz. 56 (1993) 145.

\bibitem{r9}
       V.N. Baier, V.S. Fadin, V. Khoze and E.A. Kuraev, Phys. Rep.
       78 (1981) 294.

\bibitem{r10}
       R. Barbieri, J.A. Mignaco and E. Remiddi, Il Nuovo Cimento 11A
       (1972) 824.

\bibitem{r101}
       E.A. Kuraev and V.S. Fadin, Sov. J. of Nucl. Phys. 41 (1985) 466;
       Preprint INP 84-44,  Novosibirsk, 1984;

\bibitem{r11}
       E.A. Kuraev, N. P. Merenkov and V. S. Fadin, Sov. J. of Nucl. Phys. 45
       (1987) 486.

\bibitem{r12} H. Cheng and T. T. Wu 'Expanding Protons: Scattering at High
energies', MIT Press Cambridge(Ma)-London, 1986.

\bibitem{r15} L.N. Lipatov, Sov. J. of Nucl. Phys. 20 (1974) 94;
\\      G. Altarelli and G. Parisi, Nucl. Phys. B126 (1977) 298;
\\      O. Nicrosini and L. Trentadue, Phys. Lett. 196B (1987) 551.

\bibitem{r16}
        M. Skrzypek, Acta Phys. Polonica B23 (1993) 135.

\bibitem{r18} W. Beenakker and B. Pietrzyk, Phys. Lett. B296 (1992) 241;
(ibidem) B304 (1993) 366.

\end{thebibliography}
\end{document}